\documentclass[12pt]{article}

\usepackage[margin=1in]{geometry}

\usepackage{amsmath,amssymb,mathtools,bm}
\usepackage{graphicx}
\usepackage{caption}
\usepackage{hyperref}
\usepackage{enumitem}
\hypersetup{hidelinks}

\usepackage[numbers,sort&compress]{natbib}

\graphicspath{{figures/}{./}}

\newcommand{\Tr}{\mathrm{Tr}}
\newcommand{\Var}{\mathrm{Var}}

\newcommand{\Cth}{C^{\mathrm{th}}}
\newcommand{\mc}{\mathrm{mc}}
\newcommand{\Sch}{C_{\mathrm{Sch}}}

\newcommand{\Ceff}{C_{\mathrm{eff}}}
\newcommand{\Ctyp}{C_{\mathrm{ent}}^{\mathrm{typ}}}

\begin{document}

\begin{center}
{\Large\bfseries Microcanonical Energy Sharing and a Page-like Curve for the Capacity of Entanglement}\\[0.7cm]

{\bfseries Raúl Arias}{\footnote{\texttt{rarias@fisica.unlp.edu.ar}}}

Instituto de F\'isica La Plata - CONICET and 
Departamento de F\'isica, Universidad Nacional de La Plata C.C. 67, 1900, La Plata, Argentina\\[0.5cm]

\end{center}

\vspace{0.3cm}


\begin{abstract}
We study the capacity of entanglement in the microcanonical ensemble for an
effectively additive bipartite system. Using typicality and the block structure
of the microcanonical reduced state, we show that in the thermodynamic regime
the capacity is controlled by energy-sharing fluctuations and can be expressed
purely in terms of standard thermal response data of the subsystems. As an
illustration, we apply the result to a toy model consisting of a Schwarzian
``black-hole'' sector coupled to a two-dimensional CFT radiation sector. At
fixed total energy, the growth of the radiation sector forces the common
temperature to decrease, producing a smooth Page-like single-hump curve for
the capacity. We also distinguish the capacity of the reduced microcanonical
state from that of a Haar-typical pure state in the same energy shell.  A
block-Wishart analysis shows that the effective-heat-capacity result holds on
both sides of the Page crossover, while a universal folded-normal scaling
function resolves a narrow window in which the two effective Hilbert-space
dimensions become comparable. The construction is an equilibrium
microcanonical mechanism for Page-like capacity curves, rather than a complete
dynamical evaporation calculation.
\end{abstract}

\section{Introduction}

Black holes sharpen the tension between semiclassical thermality and quantum unitarity.
Horizon thermodynamics assigns an entropy and temperature to a black hole, while Hawking’s
semiclassical calculation predicts approximately thermal radiation, leading to the information puzzle
\cite{Bekenstein1973,Hawking1975}.  A concise way to package the unitarity constraint is the
\emph{Page curve}: if the joint state of ``black hole + radiation'' is pure, the radiation
entanglement entropy must rise at early times and then decrease after a characteristic Page scale
\cite{Page1993Information}.  In gravitational settings this behavior is now understood through quantum extremal
surfaces and islands, which appear as phase transitions in the dominant saddles of replica path
integrals \cite{EngelhardtWall2015,Penington2020,AlmheiriEMM2019, AlmheiriMMMZ2020,AlmheiriReplica2020,PeningtonReplica2022}.

Entanglement entropy, however, is only the first moment of the entanglement spectrum.  A natural
refinement is the \emph{capacity of entanglement} (CoE), which measures \emph{fluctuations} of the
modular Hamiltonian and therefore probes the spectrum beyond its mean. Given a reduced density
matrix $\rho_A$, the modular Hamiltonian is $K_A:=-\log\rho_A$ on
$\mathrm{supp}(\rho_A)$, and the CoE is its variance,
\begin{equation}
  C_A \equiv \mathrm{Var}_{\rho_A}(K_A)
  = \mathrm{Tr}\!\left(\rho_A K_A^2\right) -
    \left[\mathrm{Tr}\!\left(\rho_A K_A\right)\right]^2 .
  \label{eq:CoE-variance}
\end{equation}
In holography and QFT this object has a clean meaning and has been studied from several angles
\cite{deBoer2019,NakaguchiNishioka2016, Nandy2021, Arias2023}.  More recently, replica-wormhole methods have shown that the
CoE can probe saddle structures and ensemble choices in ways that are not captured by the entropy
alone \cite{Kawabata2021Replica,Kawabata2021Probing, Arias2026, Arias2026agus}.  This motivates a basic question:
\emph{what is the simplest mechanism that produces Page-like behavior for the CoE?}

The point of this work is that a Page-like CoE curve arises already from a minimal, nongravitational
input: \emph{microcanonical typicality} plus \emph{energy sharing under a global constraint}.
We consider a bipartite system $A\cup B$ with approximately additive Hamiltonian $H\simeq H_A+H_B$
and a fixed total energy shell.  In the standard regime of canonical typicality, in which $A$ is
sufficiently small compared with its complement, Haar-typical pure states in that shell have reduced
states on $A$ sharply concentrated around the reduced microcanonical state
\cite{PopescuShortWinter2006,Goldstein2006}.
Because the total energy is fixed, the subsystem energy $E_A$ fluctuates with weights controlled by
the densities of states of both subsystems; crucially, the microcanonical modular Hamiltonian inherits
its energy dependence from the \emph{complement} $B$.  As a result, the CoE is governed by
energy-sharing fluctuations, which are set by ordinary thermodynamic response data (heat capacities).
This yields a compact leading-order prediction: the CoE behaves like an \emph{effective heat capacity}
built from $A$ and $B$.

To make the discussion concrete, we apply this mechanism to a solvable toy model in which the
``black hole'' sector is a nearly-AdS$_2$ (JT/Schwarzian) system coupled to a two-dimensional CFT
radiation sector, whose equations of state are known analytically \cite{MSY2016,SaadSS2019}.
The resulting CoE as a function of a radiation size parameter exhibits a single-hump, Page-like
curve, and its peak is explicitly ensemble dependent, illustrating in a transparent setting why the
CoE can differ qualitatively from entropy-based diagnostics. For an isolated gravitational system the microcanonical ensemble is the natural choice (fixed total ADM energy).
This constraint enforces nontrivial energy sharing between the black-hole and
radiation sectors, and the reduced state inherits an energy-block structure.
The main claim here is that this kinematical feature already controls the
entanglement capacity near $n\simeq 1$. In particular, the microcanonical
capacity differs qualitatively from what one would obtain in a standard
fixed-temperature canonical preparation, where the reduced state of the
radiation sector factorizes thermally and no Page-like peak is produced.

There is a qualification which becomes important when the radiation is not the
smaller subsystem.  Trace-norm typicality by itself does not control the
fluctuations of $-\log\rho_A$ uniformly in the presence of very small
eigenvalues.  We therefore analyze a typical pure state directly.  The fixed
energy shell decomposes into energy blocks, and in each block the reduced
density matrix is a normalized Wishart matrix.  This gives a result symmetric
under $A\leftrightarrow B$, as required by purity.  Away from the Page window it
reproduces the same effective heat capacity on either side of the crossover.
Inside that window it produces a universal smoothing, while the usual
Marchenko--Pastur contribution remains subleading.  The derivation and a direct
finite-entropy check are presented in Appendix~\ref{app:typical-pure}.

The paper is organized as follows.  In Sec.~\ref{sec:typicality} we describe the reduced
microcanonical state and state precisely where canonical typicality applies.  In
Sec.~\ref{sec:microcanonical-CoE} we derive the relation between the CoE and
energy-sharing fluctuations.  Section~\ref{sec:BH-model} applies the result to
the CFT$_2$+Schwarzian model and compares microcanonical and canonical
preparations.  Appendix~\ref{app:typical-pure} derives the typical-pure-state
extension and tests the saddle approximation without expanding the
energy-sharing integral.

\section{Microcanonical typicality and the reduced state}
\label{sec:typicality}

We consider a bipartite quantum system
\(
\mathcal{H}=\mathcal{H}_A\otimes\mathcal{H}_B
\)
with an approximately additive Hamiltonian,
\begin{equation}
  H \simeq H_A\otimes \mathbf{1}_B + \mathbf{1}_A\otimes H_B ,
  \label{eq:additive_H_s2}
\end{equation}
and we focus on states with a fixed total energy.  Concretely, we pick an energy window
\([E-\Delta/2,\,E+\Delta/2]\) and define the microcanonical projector and its dimension,
\begin{equation}
  \Pi_{\Delta}(E)\equiv
  \sum_{E_\alpha\in[E-\Delta/2,E+\Delta/2]} |\alpha\rangle\langle \alpha| ,
  \qquad
  d_{\Delta}(E)\equiv \Tr \Pi_{\Delta}(E),
  \label{eq:micro_shell_projector_s2}
\end{equation}
where \(\{|\alpha\rangle\}\) are eigenstates of \(H\).  The maximally mixed state on the shell is
$\Omega_{\Delta}(E)\equiv \frac{\Pi_{\Delta}(E)}{d_{\Delta}(E)}$,
and the reduced microcanonical state on subsystem \(A\) is
$\omega_A^{\mc}(E)\equiv \Tr_B \,\Omega_{\Delta}(E)$.

A crucial simplification is that $\omega_A^{\rm mc}(E)$ is not merely a
convenient mixed state: in the standard regime of canonical typicality it
captures the reduced density matrix of typical pure states in the energy shell.
If $|\psi\rangle$ is drawn Haar-randomly from the microcanonical subspace
$\mathcal H_\Delta(E)=\Pi_\Delta(E)\mathcal H$, then, for large shell dimension
and for subsystems $A$ that are not too large compared with their complement,
the reduced state
\begin{equation}
\rho_A^\psi=\Tr_B |\psi\rangle\langle \psi|
\label{eq:rho_typical_s2}
\end{equation}
is sharply concentrated around $\omega_A^{\rm mc}(E)$. This is the content of
microcanonical/canonical typicality~\cite{PopescuShortWinter2006,Goldstein2006}. In what
follows we first compute the capacity $C_{\rm ent}^{\mc}(A)$ of
$\omega_A^{\rm mc}$.  This is the controlled leading description of
$C(\rho_A^\psi)$ while $A$ is the smaller effective subsystem.  It should not
be used as a trace-norm argument when $A$ becomes larger than $B$, because the
logarithm in \eqref{eq:CoE-variance} is sensitive to small eigenvalues.  The
quantity $C_{\rm ent}^{\rm typ}(A):=C(\rho_A^\psi)$ is treated directly in
Appendix~\ref{app:typical-pure}, where purity supplies the continuation to the
opposite regime and also resolves the crossover.

The key structural fact is that, under \eqref{eq:additive_H_s2}, \(\omega_A^{\mc}(E)\) is (approximately)
block diagonal in the energy eigenbasis of \(H_A\).  Let \(\Pi_a^A\) be the projector onto the
\(H_A\)-eigenspace with energy \(E_a\).  Then the weight of the block \(a\) is controlled by the
number of available states of the complement \(B\) at the \emph{complementary} energy \(E-E_a\)\footnote{
Strictly speaking, $w_a^{\rm mc}$ denotes the total probability of the
$A$-energy block. The eigenvalue of $\omega_A^{\rm mc}$ for each state inside
that block is proportional to $\Omega_B(E-E_a,\Delta)$; the factor
$\Tr\Pi_a^A$ contributes to the probability distribution of $E_A$, but not to
the modular eigenvalue within the block.
}.
The microcanonical block weight is
\begin{equation}
  w_a^{\mc}
  = \frac{1}{d_{\Delta}}\,\Omega_B(E-E_a,\Delta)\,\Tr \Pi_a^A ,
  \label{eq:w_mc_s2}
\end{equation}

We use the smooth microcanonical entropy defined in
\eqref{eq:Phi_def_s3}.  Then the energy dependence of the modular Hamiltonian
is particularly simple. Because \(\omega_A^{\mc}\) is diagonal in energy
blocks, its logarithm is also block diagonal.
Up to an additive constant (fixed by normalization), the microcanonical modular Hamiltonian evaluated
on an \(A\)-energy block \(E_a\) is
\begin{equation}
  K_A^{\mc}(E_a)
  = \mathrm{const} - \log \Omega_B(E-E_a,\Delta)
  = \mathrm{const} - S_B(E-E_a).
  \label{eq:K_mc_s2}
\end{equation}
In words this is saying that \emph{the modular weight of an $A$-energy sector is the
microcanonical entropy of the complement evaluated at the complementary energy.}

Summarizing, in a fixed-energy shell the subsystem energy $E_A$ is not fixed
but fluctuates due to energy sharing with $B$.  The energy weights of
$\omega_A^{\mc}$ are controlled by the complementary density of states in
\eqref{eq:w_mc_s2}, and its modular Hamiltonian is given by
\eqref{eq:K_mc_s2}.  Therefore its modular fluctuations are driven by
energy-sharing fluctuations.  Canonical typicality identifies this result with
the capacity of \eqref{eq:rho_typical_s2} only in its usual small-subsystem
regime; the block-Wishart analysis in Appendix~\ref{app:typical-pure} gives the
extension beyond it.

\section{From energy sharing to CoE}
\label{sec:microcanonical-CoE}

We now derive the central thermodynamic result: the capacity of the reduced
microcanonical state is controlled by \emph{energy-sharing fluctuations}
between $A$ and $B$ and takes a universal effective-heat-capacity form evaluated
at the common equilibrium temperature.  Appendix~\ref{app:typical-pure} shows
when the same expression describes the typical pure state
\eqref{eq:rho_typical_s2}.

\subsection*{Energy-sharing distribution and the equilibrium saddle}
In the thermodynamic regime we may treat the subsystem energy \(E_A\) as a continuous variable.
The microcanonical weights imply that \(E_A\) is distributed according to the product of densities
of states,
\begin{equation}
  p(E_A)\ \propto\ \Omega_A(E_A,\Delta)\,\Omega_B(E-E_A,\Delta)
  \;=\;\exp\!\big(\Phi(E_A)\big),
  \label{eq:p_energy_sharing_s3}
\end{equation}
with exponent
\begin{equation}
  \Phi(E_A)\ :=\ S_A(E_A,\Delta)+S_B(E-E_A,\Delta),
  \qquad S_X(E,\Delta):=\log\Omega_X(E,\Delta).
  \label{eq:Phi_def_s3}
\end{equation}
The most probable split \(E_A^\star\) is fixed by the saddle
condition
\begin{equation}
  0=\Phi'(E_A^\star)
  \quad\Longleftrightarrow\quad
  \beta_A(E_A^\star)=\beta_B(E_B^\star),
  \qquad
  \beta_X(E):=\partial_E S_X(E),
  \qquad
  E_B^\star:=E-E_A^\star,
  \label{eq:beta_match_s3}
\end{equation}
which is the common-temperature equilibrium condition.
The curvature can be expressed entirely in terms of the thermal response.  In
fact,
\begin{equation}
 S_X''(E_X)=\frac{d\beta_X}{dE_X}
 =-\frac{1}{T^2\Cth_X},
 \qquad
 \Phi''(E_A^\star)
 =-\frac{1}{T^2}\left(\frac{1}{\Cth_A}+\frac{1}{\Cth_B}\right).
 \label{eq:Phi_curvature_s3}
\end{equation}
Writing $\delta E_A=E_A-E_A^\star$, the quadratic saddle therefore gives
\begin{equation}
 p(E_A)\simeq
 \frac{1}{\sqrt{2\pi\sigma_E^2}}
 \exp\!\left[-\frac{\delta E_A^2}{2\sigma_E^2}\right],
 \qquad
 \sigma_E^2=-\frac{1}{\Phi''(E_A^\star)}.
 \label{eq:p_gaussian_s3}
\end{equation}
The saddle is stable when the curvature in \eqref{eq:Phi_curvature_s3} is
negative.  Substituting that curvature into \eqref{eq:p_gaussian_s3} gives
\begin{equation}
  \Var(E_A)
  = T^2\,\frac{\Cth_A\,\Cth_B}{\Cth_A+\Cth_B},
  \label{eq:VarEA_heatcap_s3}
\end{equation}
where \(T=\beta^{-1}\) is the common equilibrium temperature at the saddle, and
\(\Cth_X\equiv \frac{dE_X}{dT}\) denotes the usual thermal heat capacity of subsystem \(X\) evaluated at
that temperature.  Thus the width of the energy-sharing distribution is fixed
by the combined response of the two subsystems, as in the standard
finite-reservoir fluctuation formula \cite{Callen1985,PathriaBeale2011}.

\subsection*{Linearizing the microcanonical modular Hamiltonian}
We now translate energy fluctuations into modular Hamiltonian fluctuations.
From Sec.~\ref{sec:typicality}, the microcanonical modular Hamiltonian depends on \(E_A\) through \eqref{eq:K_mc_s2}.
Expanding \(S_B(E-E_A)\) around the equilibrium split \(E_B^\star\) immediately yields the linearized
form,
\begin{equation}
  K_A^{\mc}(E_A)\ \simeq\ K^\star + \beta\,(E_A-E_A^\star),
  \label{eq:K_linear_s3}
\end{equation}
where \(K^\star\) is an irrelevant constant shift.  The interpretation is direct:
at equilibrium, \(\partial_{E_A}K_A^{\mc}=\beta\), so a small energy transfer \(\delta E_A\) produces a
modular ``work'' \(\beta\,\delta E_A\).

Taking the variance of \eqref{eq:K_linear_s3} gives the leading relation,
\begin{equation}
  C_{\rm ent}^{\mc}(A)
  =\Var_{\omega_A^{\mc}}\!\big(K_A^{\mc}\big)
  \ \simeq\ \beta^2\,\Var(E_A).
  \label{eq:Cent_beta2Var_s3}
\end{equation}
Combining \eqref{eq:Cent_beta2Var_s3} with \eqref{eq:VarEA_heatcap_s3} yields the promised universal
effective heat-capacity formula at the common equilibrium temperature:
\begin{equation}
  C_{\rm ent}^{\mc}(A)\ \simeq\
  \Ceff,
  \qquad
  \Ceff:=\frac{\Cth_A\,\Cth_B}{\Cth_A+\Cth_B}.
  \label{eq:Cent_harmonic_mean_box_s3}
\end{equation}

This harmonic-mean structure makes the qualitative shape of a Page-like curve almost inevitable once we introduce a control parameter that changes the relative sizes of the subsystems.
If the radiation sector \(A\) grows (so \(\Cth_A\) increases) while the black hole sector \(B\) effectively
shrinks (so \(\Cth_B\) decreases), then: for early stages, \(\Cth_A\ll\Cth_B\) and the right-hand side of \eqref{eq:Cent_harmonic_mean_box_s3} approaches \(\Cth_A\); at intermediate stages the two responses are comparable and the capacity is largest; at late stages, \(\Cth_B\ll\Cth_A\), and it approaches \(\Cth_B\).
This is analogous to the Page-curve logic, now applied to the \emph{capacity} rather than to the entropy:
the CoE is maximized when the two subsystems are thermodynamically comparable.

In Sec.~\ref{sec:BH-model} we make this explicit in a solvable toy model where both heat capacities are
known analytically, allowing us to plot the resulting capacity and to demonstrate its ensemble
dependence.

\section{A solvable Page-like curve and its ensemble dependence}
\label{sec:BH-model}

We now translate the universal result of Sec.~\ref{sec:microcanonical-CoE} into an explicit, closed-form
curve by choosing equations of state for the two subsystems. What we want is a
one-parameter thermodynamic model in which the radiation sector effectively
grows while the total energy remains fixed. This will allow us to see how a
single-hump, Page-like behavior for the CoE follows directly from
microcanonical energy sharing. We also compare with a standard fixed-temperature
canonical preparation, in which the reduced radiation state is thermal and the
capacity does not exhibit a Page-like peak \cite{MSY2016,SaadSS2019}.
The parameter $L$ labels a family of equilibrium systems, or equivalently a
family of equilibrium states with different numbers of accessible radiation
degrees of freedom.  It is not an evaporation time.  Accordingly, the curve
below isolates a thermodynamic mechanism with the shape of a Page curve; it is
not by itself a dynamical model of black-hole evaporation.

\subsection*{Thermodynamic model}
We identify subsystem \(A\equiv R\) with a \(1{+}1\) CFT of central charge \(c\), at temperature \(T\),
with an effective size parameter \(L\) (the amount of radiation degrees of freedom accessible/collected).
In the Cardy/high-temperature regime, the thermodynamics takes the familiar form
\begin{equation}
  E_R(T)=\frac{\pi c}{6}LT^2,\qquad
  S_R(T)=\frac{\pi c}{3}LT,\qquad
  \Cth_R(T)=\frac{dE_R}{dT}=\frac{\pi c}{3}LT.
  \label{eq:CFT_thermo_s4}
\end{equation}
Subsystem \(B\equiv BH\) is the nearly-AdS$_2$ (JT/Schwarzian) sector, whose near-extremal equation of
state is (see e.g.\ \cite{Almheiri:2014cka, MSY2016}.)
\begin{equation}
  E_{BH}^{\mathrm{exc}}(T)=\pi^2\Sch T^2,\qquad
  S_{BH}(T)=S_0+2\pi^2\Sch T,\qquad
  \Cth_{BH}(T)=\frac{dE_{BH}^{\mathrm{exc}}}{dT}=2\pi^2\Sch T.
  \label{eq:JT_thermo_s4}
\end{equation}
Here $C_{\rm Sch}$ is the Schwarzian coupling, which sets the near-extremal thermal response of the nearly-AdS$_2$ black-hole sector.

In the microcanonical setting the total \emph{excitation} energy is fixed,
\begin{equation}
  E := E_R(T)+E_{BH}^{\mathrm{exc}}(T)=\mathrm{const}.
  \label{eq:micro_constraint_s4}
\end{equation}
Substituting \eqref{eq:CFT_thermo_s4} and \eqref{eq:JT_thermo_s4} and solving for \(T\) yields
\begin{equation}
  T(L)=\sqrt{\frac{E}{\frac{\pi c}{6}L+\pi^2\Sch}}.
  \label{eq:T_of_L_s4}
\end{equation}
This simple formula already captures the microcanonical ``backreaction'': increasing the effective
radiation size \(L\) forces the equilibrium temperature to \emph{decrease} at fixed total energy.

\subsection*{Explicit CoE curve and its maximum}
\begin{figure}[!t]
  \centering
  \includegraphics[width=0.90\linewidth]{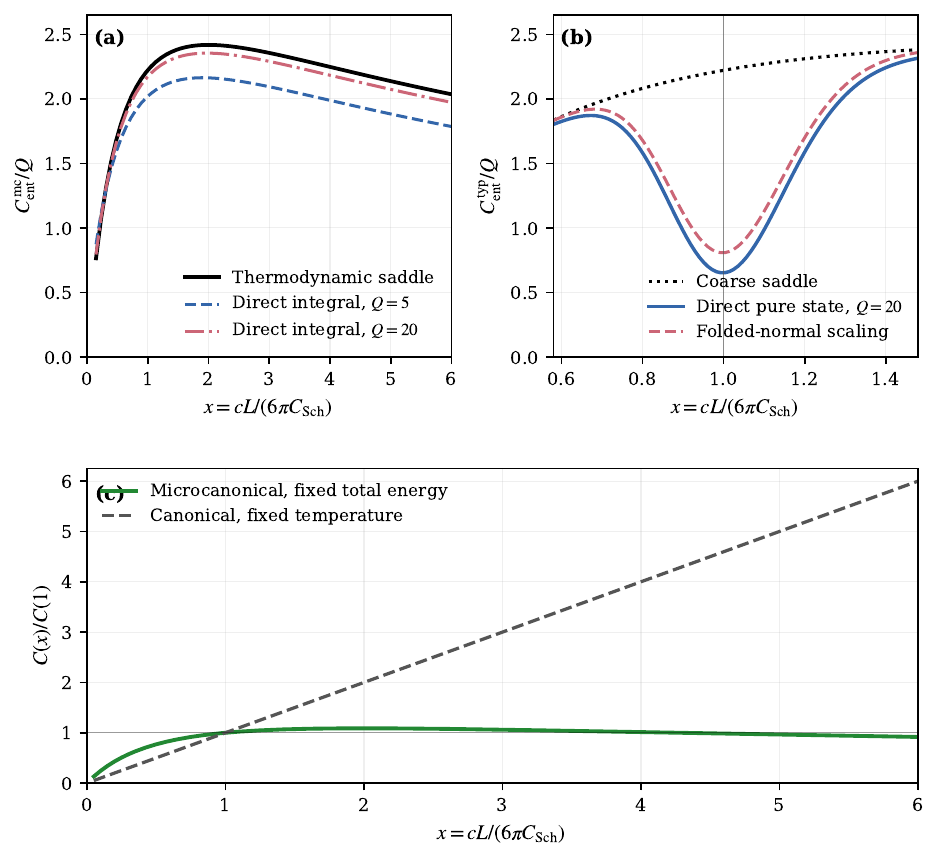}
  \caption{
  \textbf{Microcanonical Page-like capacity, finite-entropy check, and
  ensemble comparison.}
  We use $Q=\sqrt{E C_{\rm Sch}}$.  Panel (a) compares the saddle result
  \eqref{eq:universal_curve_s4} with the unexpanded density-of-states integral
  \eqref{eq:DOS_coarse_capacity_app} at $Q=5$ and $Q=20$.  Panel (b) compares
  the typical-pure-state integral \eqref{eq:DOS_typical_capacity_app} at
  $Q=20$ with the uniform approximation \eqref{eq:Ctyp_uniform_s4} and the
  coarse saddle.  Panels (a) and (b) use $S_0=0$.  Panel (c) contrasts the
  microcanonical curve with the fixed-temperature canonical result
  \eqref{eq:canonical_capacity_s4}; both are normalized by their value at
  $x=1$ to compare their shapes.
  }
  \label{fig:capacity-comparison}
\end{figure}

We now evaluate \eqref{eq:Cent_harmonic_mean_box_s3} with the responses in
\eqref{eq:CFT_thermo_s4} and \eqref{eq:JT_thermo_s4}.  Introducing
$x\equiv cL/(6\pi\Sch)$ and eliminating $T$ with \eqref{eq:T_of_L_s4} gives
\begin{equation}
  \frac{C_{\rm ent}^{\mc}(L)}{\sqrt{E\Sch}}\simeq
 2\pi\,\frac{x}{(1+x)^{3/2}}.
  \label{eq:universal_curve_s4}
\end{equation}
The curve has a single maximum at \(x=2\),
\begin{equation}
  L_{\mathrm{peak}}=\frac{12\pi}{c}\,\Sch.
  \label{eq:Lpeak_s4}
\end{equation}
Conceptually, the single-hump behavior is produced by the combination of the finite-reservoir
harmonic-mean structure and the microcanonical cooling $T(L)$. The maximum occurs when
the two heat capacities are of the same order; in the present parametrization
$C_R^{\rm th}/C_{\rm BH}^{\rm th}=x$, so the peak at $x=2$ corresponds to
$C_R^{\rm th}=2C_{\rm BH}^{\rm th}$ rather than to exact equality of the heat
capacities. The resulting dimensionless curve is shown in
Fig.~\ref{fig:capacity-comparison}(a).

In this toy model an entropy-balancing scale{\footnote{
For clarity, our Page-like point is defined by equating the \emph{thermal} parts of the
entropies, i.e.\ after subtracting the constant extremal term $S_0$, which does not
affect fluctuations in the regime considered.}} is naturally defined by balancing the radiation entropy
against the \emph{thermal} part of the black-hole entropy (i.e.\ subtracting \(S_0\)).  This gives
\begin{equation}
  L_{\mathrm{Page}}=\frac{6\pi}{c}\,\Sch,
  \qquad
  L_{\mathrm{peak}}=2\,L_{\mathrm{Page}}.
  \label{eq:Lpage_relation_s4}
\end{equation}
The take-away is not the factor of $2$ itself, but the qualitative point: the
CoE peak need not coincide with the entropy-balancing scale. Including $S_0$ would shift the
entropy-based Page scale, while leaving the fluctuation mechanism for the CoE
unchanged.

\paragraph{Typical pure states and the Page window.}

The curve \eqref{eq:universal_curve_s4} is the capacity of the reduced
microcanonical state.  To compare it with a Haar-typical pure state, the
relevant measure of the two effective block dimensions is the entropy
difference at the equilibrium saddle.  Equations \eqref{eq:CFT_thermo_s4},
\eqref{eq:JT_thermo_s4}, and \eqref{eq:T_of_L_s4} give
\begin{equation}
 \Delta_\star(L)
 :=S_R(T(L))-S_{BH}(T(L))
 =2\pi\sqrt{E\Sch}\,\frac{x-1}{\sqrt{1+x}}-S_0.
 \label{eq:Delta_star_s4}
\end{equation}
The crossover at which the two effective Hilbert-space dimensions agree is
defined by $\Delta_\star=0$.  It need not coincide with the thermal
entropy-balancing point in \eqref{eq:Lpage_relation_s4}, because the former is
sensitive to $S_0$.

The block-Wishart calculation in Appendix~\ref{app:typical-pure} gives a
uniform result in the scaling window $\Delta_\star=O(\sqrt{\Ceff})$,
\begin{equation}
 \Ctyp(L)
 =\Ceff(L)\,\mathcal F\!\left(u(L)\right)+o(\Ceff),
 \qquad
 u(L):=\frac{\Delta_\star(L)}{2\sqrt{\Ceff(L)}},
 \label{eq:Ctyp_uniform_s4}
\end{equation}
where $\Ceff(L)$ is defined by \eqref{eq:Cent_harmonic_mean_box_s3} and
\begin{equation}
 \mathcal F(u)
 =1+u^2-
 \left[
  \sqrt{\frac{2}{\pi}}e^{-u^2/2}
  +u\,\operatorname{erf}\!\left(\frac{u}{\sqrt{2}}\right)
 \right]^2.
 \label{eq:folded_scaling_s4}
\end{equation}
For $|u|\gg1$, the scaling function tends to one and
\eqref{eq:universal_curve_s4} also gives the typical-pure-state capacity on
either side of the crossover.  At $u=0$, it is reduced by the universal factor
$\mathcal F(0)=1-2/\pi$.  Thus the single-hump thermodynamic mechanism
survives, but a narrow Page window is resolved by the purity constraint.  The
finite-entropy density-of-states calculation in
Appendix~\ref{app:finite-check} tests both limits directly.
\paragraph{Relation to replica-wormhole capacity peaks.}

The microcanonical EOW-brane model of
Ref.~\cite{Kawabata2021Probing} provides a useful comparison.  There the
reduced radiation state is described, in the planar limit, by a single
Wishart matrix whose aspect ratio is fixed by the ratio between the radiation
and black-hole Hilbert-space dimensions.  Its capacity is therefore the
$O(1)$ Marchenko--Pastur function appearing in
\eqref{eq:MP_capacity_app}: it is maximal when the two dimensions agree and
is interpreted as the smooth planar resolution of the competition between
disconnected and connected replica-wormhole topologies.  In our construction
the same random-matrix contribution is present, but only within each energy
block in \eqref{eq:block_capacity_identity_app}.

The leading effect here instead comes from the second term of
\eqref{eq:block_capacity_identity_app}, namely the fluctuations of the mean
modular energy among blocks with different energy splits.  It is of order
$\Ceff$ and produces the thermodynamic curve
\eqref{eq:universal_curve_s4}, whose maximum
\eqref{eq:Lpeak_s4} need not coincide with the equal-dimension crossover
defined by \eqref{eq:Delta_star_s4}.  Near the latter, purity gives the
folded-normal scaling \eqref{eq:Ctyp_uniform_s4}, whereas the
Marchenko--Pastur term remains below leading thermodynamic order.  No change
of a gravitational replica saddle is invoked in this derivation: the hump
follows from the fixed-energy constraint and the resulting microcanonical
cooling.  The two mechanisms are therefore complementary, but physically
distinct.  In particular, a Page-like maximum of the CoE is not by itself a
unique signature of a replica-wormhole transition; its scaling with the
thermodynamic size and its location relative to the entropy crossover
distinguish the two effects.

\paragraph{Comparison with a canonical preparation.}

A central message of the toy model is that the hump relies on the
microcanonical fixed-energy constraint. In the microcanonical calculation, the
radiation and black-hole energies are not independent: if the radiation carries
energy $E_R$, the black-hole sector carries $E-E_R$. This finite-reservoir
constraint is what leads to \eqref{eq:Cent_harmonic_mean_box_s3}.

By contrast, in a standard canonical preparation at fixed temperature, with an
approximately additive Hamiltonian,
\begin{equation}
\rho_{R\,{\rm BH}}^{\rm can}
=
\frac{e^{-\beta(H_R+H_{\rm BH})}}{Z_R Z_{\rm BH}}
=
\rho_R^{\rm can}\otimes \rho_{\rm BH}^{\rm can},
\qquad
\rho_R^{\rm can}
=
\frac{e^{-\beta H_R}}{Z_R}.
\label{eq:canonical_factorized_s4}
\end{equation}
The modular Hamiltonian of the radiation sector is then simply
\begin{equation}
K_R^{\rm can}
=
-\log \rho_R^{\rm can}
=
\beta H_R+\log Z_R ,
\label{eq:canonical_modular_s4}
\end{equation}
and therefore
\begin{equation}
C_{\rm ent,R}^{\rm can}
=
\Var_{\rho_R^{\rm can}}(K_R^{\rm can})
=
\beta^2 \Var_{\rho_R^{\rm can}}(H_R)
=
\Cth_R(T)
=
\frac{\pi c}{3}LT .
\label{eq:canonical_capacity_s4}
\end{equation}
Thus the standard canonical result grows linearly with the radiation size at
fixed temperature and has no intermediate maximum. The Page-like peak is therefore tied to the fixed-total-energy
microcanonical constraint, together with the induced cooling $T(L)$.

\section{Conclusions}
\label{sec:conclusions}

We presented a compact thermodynamic mechanism---microcanonical energy sharing
plus typicality---that produces Page-like behavior for the capacity of
entanglement without a bulk replica calculation.  In an approximately additive
system, the global energy constraint produces the distribution
\eqref{eq:p_energy_sharing_s3}, whose variance is fixed by the ordinary thermal
responses through \eqref{eq:VarEA_heatcap_s3}.  The energy dependence of the
microcanonical modular Hamiltonian is inherited from the complementary density
of states, as shown in \eqref{eq:K_mc_s2}.  These facts lead to the effective
heat capacity in \eqref{eq:Cent_harmonic_mean_box_s3}.  It is small when either
subsystem is thermodynamically stiff and largest when their responses are
comparable.

Specializing to a solvable toy model---a two-dimensional CFT radiation sector
coupled to a nearly-AdS$_2$ Schwarzian black-hole sector~\cite{MSY2016, SaadSS2019, Almheiri:2014cka}---makes the discussion concrete. The microcanonical
constraint forces the common equilibrium temperature to vary with the radiation
size parameter, producing the explicit single-hump curve
\eqref{eq:universal_curve_s4}.
In this model the CoE peak occurs at a scale that need not coincide with the
entropy-balancing point, highlighting that the CoE probes a different aspect of
the entanglement spectrum. By contrast, in a standard fixed-temperature
canonical preparation the reduced radiation state is thermal and its capacity
is given by \eqref{eq:canonical_capacity_s4}, so no Page-like maximum is
produced. The peak is therefore a genuinely microcanonical finite-reservoir
effect.  The parameter $L$ labels this family of equilibrium preparations and
must not be interpreted as an evaporation time.

In the JT/Schwarzian toy model the black-hole heat capacity is set by the
Schwarzian coupling, while the radiation contribution is controlled by the CFT
central charge and the effective size $L$. The maximum determined by
\eqref{eq:Lpeak_s4} can therefore be read as the
point where the radiation and black-hole thermal responses are of the same
order, with the precise location fixed by the interplay between the
finite-reservoir harmonic mean and the microcanonical cooling $T(L)$. In the
present model the radiation response is twice the black-hole response at the
peak.  The thermal entropy-balancing scale is instead fixed by
\eqref{eq:Lpage_relation_s4}, showing explicitly that the CoE peak and the
entropy scale need not coincide.

We also separated two statements that are often conflated.  Canonical
typicality controls the reduced state only while the observed subsystem is
sufficiently small.  The direct block-Wishart calculation avoids applying that
statement to $-\log\rho_A$ outside its regime of validity.  It proves that the
effective-heat-capacity result continues on the other side by purity, while the
uniform expression \eqref{eq:Ctyp_uniform_s4} resolves the window in which the
two block dimensions are comparable.  Marchenko--Pastur fluctuations within an
energy block contribute only below the leading thermodynamic order.  The direct
density-of-states evaluation confirms the saddle formula as the entropy is
increased and displays the predicted Page-window suppression.

The take-home message is simple: energy conservation plus typicality imposes a
rigid and testable structure on entanglement-spectrum fluctuations. In a
fixed-energy shell, the modular Hamiltonian of a subsystem inherits its leading
energy dependence from the density of states of the complement, and the CoE is
controlled by ordinary energy-sharing fluctuations. This perspective
complements fully gravitational replica computations by isolating a
thermodynamic mechanism for Page-like capacity curves and by clarifying which
features follow from the microcanonical constraint itself, which ones require
purity, and which ones would require a genuinely dynamical evaporation model.

\appendix

\section{Typical pure states in the microcanonical shell}
\label{app:typical-pure}

In the main text we first computed the capacity of the reduced
microcanonical state $\omega_A^{\mc}$.  We now compute the capacity of the
typical pure-state reduction \eqref{eq:rho_typical_s2} directly.  This avoids
using trace-norm typicality for a quantity containing $\log\rho_A$ and makes the
$A\leftrightarrow B$ symmetry imposed by purity manifest.

\subsection{Energy blocks and the Wishart ensemble}

Under the additive approximation \eqref{eq:additive_H_s2}, a sufficiently
narrow energy shell has the direct-sum decomposition
\begin{equation}
 \mathcal H_\Delta(E)
 \simeq
 \bigoplus_a
 \left(\mathcal H_{A,a}\otimes\mathcal H_{B,\bar a}\right),
 \qquad
 d_{A,a}:=\dim\mathcal H_{A,a},
 \qquad
 d_{B,a}:=\dim\mathcal H_{B,\bar a},
 \label{eq:shell_direct_sum_app}
\end{equation}
where the two factors in each summand carry energies $E_a$ and $E-E_a$,
respectively.  A Haar-random vector in \eqref{eq:shell_direct_sum_app} can be
represented by independent complex Gaussian matrices $G_a$ as
\begin{equation}
 |\psi\rangle
 =\frac{1}{\sqrt{\mathcal T}}
 \sum_a\sum_{i=1}^{d_{A,a}}\sum_{\alpha=1}^{d_{B,a}}
 (G_a)_{i\alpha}\,
 |a,i\rangle_A|\bar a,\alpha\rangle_B,
 \qquad
 \mathcal T:=\sum_b\Tr(G_bG_b^\dagger).
 \label{eq:Haar_block_state_app}
\end{equation}
Tracing over $B$ turns \eqref{eq:rho_typical_s2} into a direct sum of
normalized Wishart matrices,
\begin{equation}
 \rho_A^\psi
 =\bigoplus_a p_a\sigma_a,
 \qquad
 \sigma_a:=\frac{G_aG_a^\dagger}{t_a},
 \qquad
 p_a:=\frac{t_a}{\mathcal T},
 \qquad
 t_a:=\Tr(G_aG_a^\dagger).
 \label{eq:rho_block_Wishart_app}
\end{equation}
The variables $t_a$ are independent gamma variables with shape
$d_{A,a}d_{B,a}$, while $\sigma_a$ is independent of $t_a$ within each block
\cite{ZyczkowskiSommers2001,deBoer2019,Okuyama2021}.  Consequently the vector
of block probabilities is Dirichlet distributed.  Defining
\begin{equation}
 D:=\sum_b d_{A,b}d_{B,b},
 \qquad
 q_a:=\frac{d_{A,a}d_{B,a}}{D},
 \label{eq:q_block_app}
\end{equation}
one has
\begin{equation}
 \mathbb E(p_a)=q_a,
 \qquad
 \Var(p_a)=\frac{q_a(1-q_a)}{D+1}.
 \label{eq:Dirichlet_concentration_app}
\end{equation}
The block weights therefore concentrate around \eqref{eq:q_block_app} when the
shell dimension is large.  This concentration is stronger than the
thermodynamic corrections retained below.

\subsection{An exact decomposition of the capacity}

Consider any block-diagonal state of the form appearing in
\eqref{eq:rho_block_Wishart_app}.  Within block $a$, the modular Hamiltonian is
$-\log p_a-\log\sigma_a$.  The law of total variance applied to
\eqref{eq:CoE-variance} gives the exact identity
\begin{equation}
 C\!\left(\bigoplus_a p_a\sigma_a\right)
 =\sum_a p_a C(\sigma_a)
 +\Var_p\!\left[-\log p_a+S(\sigma_a)\right].
 \label{eq:block_capacity_identity_app}
\end{equation}
The first term measures fluctuations within the energy blocks.  The second
term measures fluctuations of their mean modular energies and is the term that
contains the leading thermodynamic contribution.

For each block define
\begin{equation}
 m_a:=\min(d_{A,a},d_{B,a}),
 \qquad
 M_a:=\max(d_{A,a},d_{B,a}),
 \qquad
 \gamma_a:=\frac{m_a}{M_a}\leq1.
 \label{eq:block_aspect_ratio_app}
\end{equation}
In the planar Wishart limit, governed by the Marchenko--Pastur law
\cite{MarchenkoPastur1967,ZyczkowskiSommers2001}, the Page correction to the
entropy and the capacity within a block are
\begin{equation}
 S(\sigma_a)=\log m_a-\frac{\gamma_a}{2}+o(1),
 \qquad
 C(\sigma_a)=\mathcal V_{\rm MP}(\gamma_a)+o(1),
 \label{eq:Wishart_entropy_capacity_app}
\end{equation}
where the Marchenko--Pastur capacity is
\begin{equation}
 \mathcal V_{\rm MP}(\gamma)
 =-1-\frac{3\gamma}{2}-\frac{\gamma^2}{4}
 +\left(\gamma-\gamma^{-1}\right)\log(1-\gamma)
 +2\operatorname{Li}_2(\gamma).
 \label{eq:MP_capacity_app}
\end{equation}
The expression in \eqref{eq:MP_capacity_app} is understood by continuity at
$\gamma=1$, where it equals $\pi^2/3-11/4$; it vanishes as $\gamma\to0$.
Finite-dimensional expressions are given in Ref.~\cite{Okuyama2021}.

Using the concentration in \eqref{eq:Dirichlet_concentration_app}, the mean
modular energy of block $a$ becomes
\begin{equation}
 -\log q_a+S(\sigma_a)
 =\log D-\log M_a-\frac{\gamma_a}{2}+o(1).
 \label{eq:block_mean_modular_app}
\end{equation}
Combining \eqref{eq:block_capacity_identity_app},
\eqref{eq:Wishart_entropy_capacity_app}, and
\eqref{eq:block_mean_modular_app} gives
\begin{equation}
 \Ctyp
 =\Var_q\!\left[
    \log D-\log M_a-\frac{\gamma_a}{2}
  \right]
 +\sum_a q_a\mathcal V_{\rm MP}(\gamma_a)
 +o(\Ceff).
 \label{eq:Ctyp_master_app}
\end{equation}
Unlike the reduced-microcanonical result, \eqref{eq:Ctyp_master_app} depends on
the larger of the two block dimensions.  It is explicitly invariant under
$A\leftrightarrow B$, as required because the two reductions of a pure state
have identical nonzero spectra.

\subsection{Thermodynamic limit away from the crossover}

Set $\delta E_A=E_A-E_A^\star$.  In the energy range selected by
\eqref{eq:p_gaussian_s3}, suppose first that one block dimension is
parametrically larger than the other.  If $d_{B}\gg d_A$, then $M=d_B$ in
\eqref{eq:block_mean_modular_app} and its linear fluctuation is
$+\beta\delta E_A$.  If $d_A\gg d_B$, then $M=d_A$ and the linear fluctuation
is $-\beta\delta E_A$.  The sign changes, but its variance does not.  Therefore
\eqref{eq:VarEA_heatcap_s3} and \eqref{eq:Ctyp_master_app} imply
\begin{equation}
 \Ctyp=\Ceff+O(1)
 \label{eq:Ctyp_away_app}
\end{equation}
on either side of the crossover.  In this regime $\gamma$ is exponentially
small in the entropy difference, so the Marchenko--Pastur term in
\eqref{eq:Ctyp_master_app} is also below the leading thermodynamic order.  This
proves directly, without extending canonical typicality beyond the smaller
subsystem, that the harmonic-mean result continues to the opposite side by
purity.

\subsection{Uniform Page-window scaling}

The approximation used to obtain \eqref{eq:Ctyp_away_app} is not uniform when
the two dimensions become comparable within the energy fluctuation window.
Introduce
\begin{equation}
 \Delta(E_A):=S_A(E_A)-S_B(E-E_A),
 \qquad
 \Delta(E_A)=\Delta_\star+2\beta\delta E_A+O(1),
 \label{eq:entropy_difference_app}
\end{equation}
where $\Delta_\star$ is evaluated at the saddle.  Since
$\log M=(S_A+S_B+|\Delta|)/2$ and the total entropy has no linear variation at
the saddle, the leading fluctuating part of \eqref{eq:block_mean_modular_app}
is
\begin{equation}
 \overline K_a-\mathrm{const}
 =-\frac{1}{2}\left|\Delta_\star+2\beta\delta E_A\right|+O(1).
 \label{eq:folded_modular_app}
\end{equation}
Equation \eqref{eq:VarEA_heatcap_s3} allows us to write
$\delta E_A=T\sqrt{\Ceff}\,Z$, with $Z$ a standard normal variable.  Using
$u$ defined in \eqref{eq:Ctyp_uniform_s4}, \eqref{eq:folded_modular_app}
becomes $-\sqrt{\Ceff}|Z+u|$ at leading order.  The mean of the folded normal
variable is
\begin{equation}
 \mathbb E|Z+u|
 =\sqrt{\frac{2}{\pi}}e^{-u^2/2}
 +u\,\operatorname{erf}\!\left(\frac{u}{\sqrt{2}}\right).
 \label{eq:folded_mean_app}
\end{equation}
Its variance gives precisely \eqref{eq:folded_scaling_s4}, proving the uniform
formula \eqref{eq:Ctyp_uniform_s4}.  Blocks with $\gamma=O(1)$ occupy a region
of width $O(\Ceff^{-1/2})$ in the Gaussian variable.  Hence both the explicit
Marchenko--Pastur term and the $\gamma/2$ correction in
\eqref{eq:Ctyp_master_app} are subleading compared with the
$O(\Ceff)$ folded-normal contribution.

For the CFT$_2$+Schwarzian model it is useful to make the location of the
crossover explicit.  Defining
\begin{equation}
 s:=\frac{S_0}{2\pi\sqrt{E\Sch}},
 \qquad
 x_{\rm cross}
 =\left(\frac{s+\sqrt{s^2+8}}{2}\right)^2-1,
 \label{eq:x_cross_app}
\end{equation}
one verifies from \eqref{eq:Delta_star_s4} that
$\Delta_\star(x_{\rm cross})=0$.  For $S_0=0$ this gives $x_{\rm cross}=1$.
The condition $|\Delta_\star|=O(\sqrt{\Ceff})$ implies a width
$\Delta x=O((E\Sch)^{-1/4})$.  The Page-window correction is therefore narrow
in the large-entropy limit, unless \eqref{eq:x_cross_app} is tuned close to the
maximum in \eqref{eq:Lpeak_s4}.

\section{A direct finite-entropy density-of-states check}
\label{app:finite-check}

We finally check the saddle approximation without expanding the
energy-sharing integral.  This is a finite-entropy test within the leading
Cardy and Schwarzian densities of states; it does not include theory-dependent
subleading corrections to those densities.  Let
\begin{equation}
 y:=\frac{E_R}{E},
 \qquad
 Q:=\sqrt{E\Sch}.
 \label{eq:y_Q_app}
\end{equation}
Eliminating the temperature from \eqref{eq:CFT_thermo_s4} and
\eqref{eq:JT_thermo_s4} gives the microcanonical entropies
\begin{equation}
 S_R(y)=2\pi Q\sqrt{xy},
 \qquad
 S_{BH}(y)=S_0+2\pi Q\sqrt{1-y}.
 \label{eq:DOS_entropies_app}
\end{equation}
The normalized energy-sharing density is then
\begin{equation}
 P_x(y)
 :=\frac{\exp[S_R(y)+S_{BH}(y)]}
 {\displaystyle\int_0^1 dz\,\exp[S_R(z)+S_{BH}(z)]}.
 \label{eq:DOS_probability_app}
\end{equation}
Since the additive constant in the modular Hamiltonian drops out of a
variance, the direct coarse-grained result is
\begin{equation}
 C_{\rm DOS}^{\mc}(x,Q)
 :=\Var_{P_x}\!\left[S_{BH}(y)\right].
 \label{eq:DOS_coarse_capacity_app}
\end{equation}
No Gaussian approximation has been made in
\eqref{eq:DOS_coarse_capacity_app}.  Its large-$Q$ saddle is
\eqref{eq:universal_curve_s4}.

The corresponding large-block typical-pure-state integral follows directly
from \eqref{eq:Ctyp_master_app}.  With
$\gamma(y)=\exp[-|S_R(y)-S_{BH}(y)|]$, it is
\begin{align}
 C_{\rm DOS}^{\rm typ}(x,Q)
 &:={\Var}_{P_x}\!\left[
  -\max(S_R(y),S_{BH}(y))-\frac{\gamma(y)}{2}
 \right]
 +\int_0^1dy\,P_x(y)\,\mathcal V_{\rm MP}(\gamma(y)).
 \label{eq:DOS_typical_capacity_app}
\end{align}
Figure~\ref{fig:capacity-comparison}(a) shows that
\eqref{eq:DOS_coarse_capacity_app} approaches the saddle curve as $Q$ grows.
Panel (b) evaluates \eqref{eq:DOS_typical_capacity_app}; away from the
crossover it approaches the same curve, while near $x=1$ it develops the
folded-normal suppression predicted by \eqref{eq:Ctyp_uniform_s4}.  This check
also confirms that the internal Marchenko--Pastur term is subleading.

\bibliographystyle{unsrtnat}
\bibliography{microcanonical_energy_sharing}

\end{document}